\documentclass[aps, prl, 10pt, twocolumn, groupedaddress, superscriptaddress, nofootinbib]{revtex4-2}

\usepackage[dvipsnames]{xcolor} 
\usepackage{graphicx}
\usepackage{dcolumn}
\usepackage{bm}
\usepackage{multirow}

\definecolor{myblue}{RGB}{34,31,150} 
\usepackage[breaklinks=true,colorlinks=true,linkcolor=myblue,urlcolor=myblue,citecolor=myblue]{hyperref} 

\begin{document}

\preprint{APS/123-QED}

\title{\textbf{{Exploiting complex 3D-printed surface structures for portable quantum technologies}} 
}%

\author{N. Cooper}
\email{Contact author: nathan.cooper@nottingham.ac.uk}
\affiliation{School of Physics and Astronomy, University of Nottingham, University Park, Nottingham, NG7 2RD, UK}%
\author{D. Johnson}%
\affiliation{School of Physics and Astronomy, University of Nottingham, University Park, Nottingham, NG7 2RD, UK}%
\author{B. Hopton}
\affiliation{School of Physics and Astronomy, University of Nottingham, University Park, Nottingham, NG7 2RD, UK}%
\author{M. Overton}
\affiliation{School of Physics and Astronomy, University of Nottingham, University Park, Nottingham, NG7 2RD, UK}%

\author{D. Stupple}
\author{A. Bratu}
\author{E. Wilson}
\author{J. Robinson}
\affiliation{Torr Scientific Ltd, Pebsham Lane, Bexhill-on-Sea, TN40 2RZ
}%
\author{L. Coles}
\author{M. Papastavrou}
\affiliation{Metamorphic Additive Manufacturing Ltd, Riverside Chambers, Full Street
Derby, DE1 3AF, UK
}%
\author{L. Hackermueller}
\affiliation{School of Physics and Astronomy, University of Nottingham, University Park, Nottingham, NG7 2RD, UK}%

\date{\today}

\begin{abstract}

Portable quantum technologies require robust, lightweight apparatus with superior performance. For techniques dependent upon high-vacuum environments, such as atom interferometers and atomic clocks, 3D-printing enables new avenues to tailor in-vacuum gas propagation dynamics. We demonstrate intricate, fine-scale surface patterning of 3D-printed vacuum components to increase the rate at which gas particles collide with the surface. By applying a non-evaporable getter coating for use as a surface pump, we show that the patterned surface pumps gas particles 3.8 times faster than an equivalent flat areas. These patterns can be directly integrated into additively manufactured components, enabling application in close proximity to key experimental regions and contributing to overall mass-reduction. 
We develop numerical simulations that show good agreement with this result and predict up to a ten-fold increase in pumping rate, for realistic surface structures. Our work has direct applications in enabling passively-pumped portable quantum technologies, but also establishes 3D-printing as a powerful technique for the creation of optimized surface patterning to provide enhanced control over high-vacuum gas dynamics for a broad range of applications.
\end{abstract}

\maketitle

\section{Introduction}

A rapidly growing range of portable atom-based quantum technologies (QT) are exploiting ultra-high vacuum (UHV) conditions to enable sensing and timing applications with unprecedented performance \cite{gravimeter2,gravimeter1,portmag1,PRAp_mag1,portclock,PRAp_clock1,portaccelerometer}. A significant challenge for field-employment is to reduce the size, weight and power consumption (SWAP) of vacuum equipment, especially pumps. While much progress in the miniaturization of experimental hardware such as optics \cite{optamot,optamot2,compactlaser,PRAp_minilock,AM_VC}, vacuum vessels \cite{quartz_vc, AMUHV}, magnetic field generation \cite{saint_coils}, control electronics \cite{arduinolock} and integrated atom traps \cite{PRAp_minimot} has been demonstrated, vacuum pumps have developed more slowly and are now a major limiting factor. Solving this problem will facilitate the deployment of portable QT for both industry and research. In addition to terrestrial QT applications, a range of exciting proposals to deploy space-based QT~\cite{QPinSpace} have been put forward for Earth monitoring~\cite{Earth_obs1,Earth_obs2}, astronomical observations \cite{planet_obs} and precision tests of fundamental physics \cite{equivtest2,spaceint,AION}; low mass, power consumption and heat dissipation are essential for such deployments. 
Understanding and exploiting the benefits offered by 3D-printed surfaces---in particular in combination with 3D-printed chambers or hardware elements---can help overcome many of these challenges.

Passive pumping via non-evaporable getters (NEGs) \cite{NEGref, NEGref2, NEGpill} offers a promising route to 
unlock the full potential of portable QT \cite{NEGMOT, LongNEGMOT}. However, substantial increases in pumping rate and pump capacity are required, particularly for some ``problem" particle species with low pumping efficiency \cite{stickprob}.  

By considering the physics of a diffuse gas interacting with a patterned surface, it is possible to engineer complex surfaces that enhance the performance of high-vacuum systems for QT; specifically, we show that additive manufacturing (AM, also known as 3D-printing) can create intricate surface structures that substantially increase the effectiveness of passive NEG pumps. Exploiting AM, these surfaces can be embedded directly within the walls of vacuum components, aiding compactness and facilitating integration into target systems.
We demonstrate experimentally that appropriately structured parts of a 3D-printed, NEG-coated surface can pump 3.8 times faster than equivalent flat regions. 

We present numerical simulations, validated by our experimental results, which suggest up to a 10-fold increase in pumping speed with readily-achievable surface geometries.
The resulting improvement in pumping rate through complex geometries can exceed the corresponding increase in total surface area. Furthermore, this enhancement can be realized without any increase in the projected surface area---which serves as a metric for the spatial footprint of the surface pump within the vacuum system.
This approach to modeling the behavior of a rarefied gas interacting with complex surface patterning thus reveals optimal structures for NEG surface pumping, while also paving the way for AM-enabled surface texture engineering to be applied more broadly to meet the needs of emerging QT in high-vacuum environments; this could include, for example, the use of optimized surface patterns to control channel conductances, particle dwell times or nozzle output distributions.  

\begin{figure}[ht]
    \centering
    \includegraphics[width=0.5\textwidth]{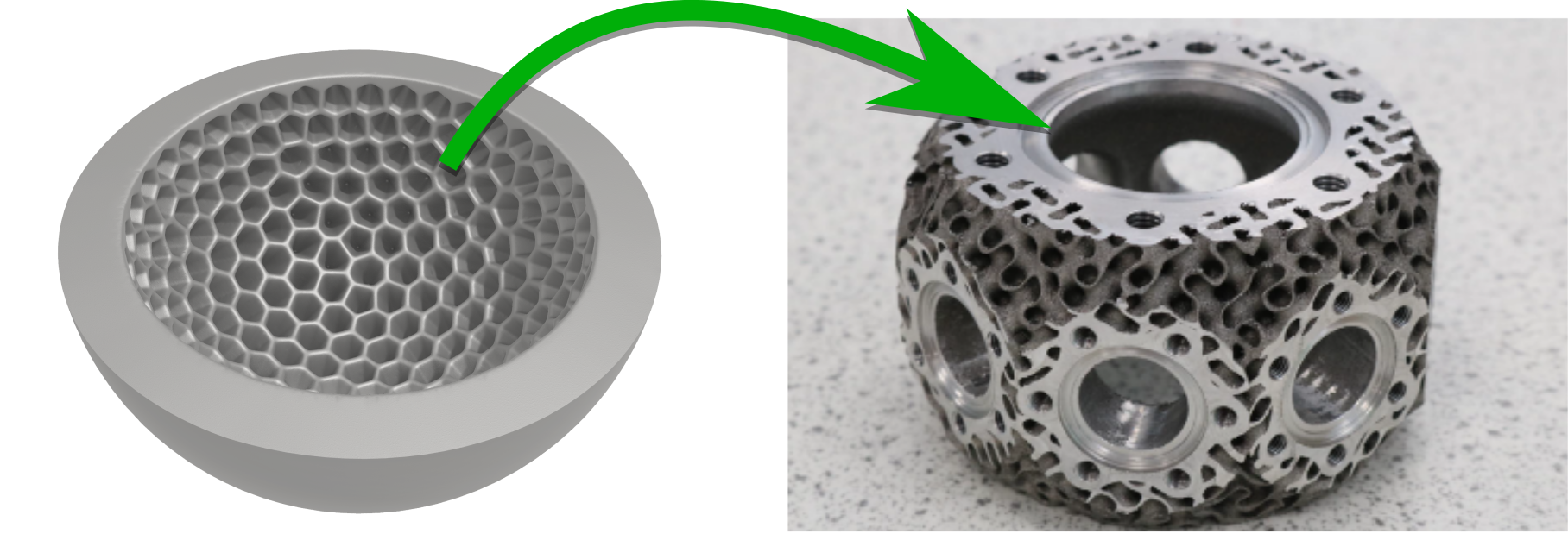}
    \caption{Additively manufactured vacuum apparatus with internal surface structuring: the surface structures could be conformally mapped onto the interior surfaces of chambers and components. Herein it is shown that, with an NEG coating applied to such a surface, this can enable large improvements in surface pump performance. Similar methods will likely enable tailored control of gas flow within high-vacuum apparatus. The AM chamber photographed here is described fully in \cite{AMUHV}.}
    \label{fig:vacuumchamber}
\end{figure}

\begin{figure*}[ht]
\centering
\includegraphics[width=0.7\linewidth]{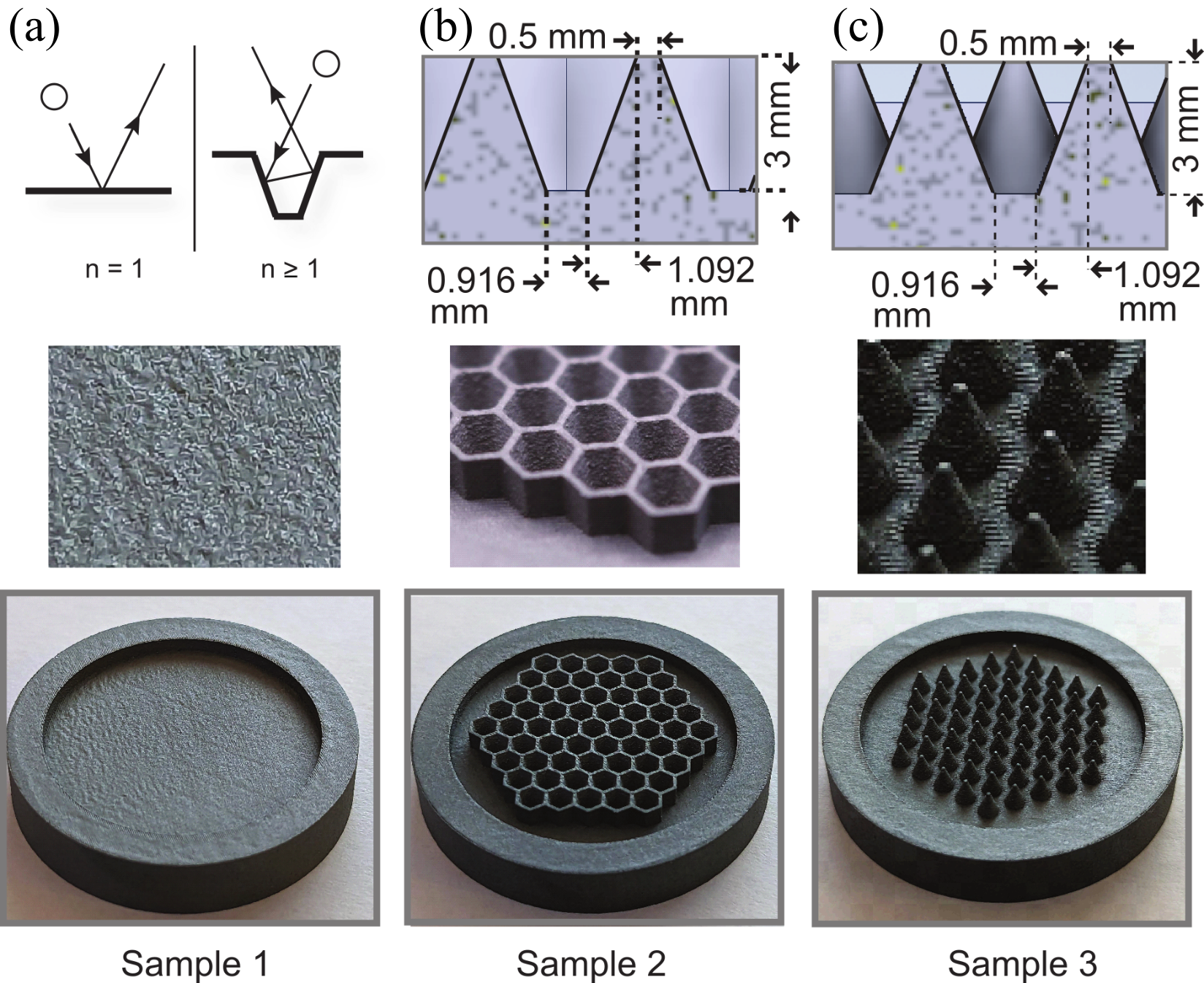}
\caption{The 3D-printed surface structures experimentally tested for pumping performance, including a flat reference surface. (a) Diagram illustrating how tailored  surface structures enable multiple collisions; photographs of flat reference surface. (b) Array of tapered, truncated hexagonal pockets. (c) Hexagonally-packed array of conical protrusions.} 
\label{samples}
\end{figure*}

We produce AM surface structures in titanium alloy via additive manufacturing and they are confirmed as UHV compatible. Photographs of the surface structures tested are shown in figure \ref{samples}(a) and their exact geometries are described below. Previous work has not only demonstrated AM of UHV components \cite{birmingham_flange}, but has also shown that AM allows conformal application of intricate structures over complex 3D surfaces \cite{AMUHV,optamot}. Combined with the approach described herein, this establishes a clear pathway to integrated, structured pumping surfaces, suitable to cover a large fraction of the internal surface area of a vacuum system, within nearly any UHV component---see figure \ref{fig:vacuumchamber}.

Beyond QT, passive surface pumps play a role in maintaining UHV environments within a wide range of technologies and devices, from electron microscopy \cite{UHVelectronmicroscopy} and mass spectrometry \cite{C8JA00087E} to prototype fusion reactors \cite{ITER} and particle accelerators \cite{CERN_NEG}. All such applications may benefit from improved pumping performance and the technology introduced here.

\section{Theory and simulations}

Monte Carlo simulation methods, based on ballistic particle propagation in the long mean-free-path limit, were used to model the interactions of incident gas particles with structured surfaces. This approach allows rapid investigation of the properties of a wide range of surface structures, enabling optimization of surface patterns for specific applications. The employment of a Monte Carlo approach based on sequential simulation of individual particles makes the simulation process highly general; any surface pattern with either a repeated unit cell can be simulated. The generality of the method facilitates analysis of complex surface geometries, allowing surface patterns to be engineered to forms optimal for specific applications.

Here we target the application of optimizing pumping performance for surfaces coated with NEG material. 
Small-scale surface patterning increases the surface area available and thus yields a corresponding increase in total pumping capacity. However, these structures can also increase the pumping rate by factors significantly larger than the increase in surface area. The pumping rate is limited by the finite probability for an incident gas particle to be absorbed, via either binding or embedding, upon collision with the surface. This probability, $P_s$, depends on the particle species and getter material, but often  $P_s \ll 1$~\cite{stickprob}.  

Complex surface structures can enhance pumping efficiency for any $P_s < 1$. To simulate particle interactions with complex surfaces, we consider particles incident upon both a flat and a structured surface, as depicted in figure \ref{samples}(a). A particle incident upon the flat surface impacts the surface only once, yielding overall pumping probability $P_s$. However, a particle incident upon the structured surface, if not bound by the first collision, has a chance to re-impact the surface \emph{before} passing back through the virtual surface marking the top-plane of the surface structure, and can undergo multiple collisions in this way. This yields an effective pumping probability $P_e > P_s$. 

To enable evaluation of $P_e$, we define $n \geq 1$ as the number of surface collisions that would be undergone by an incident, non-sticking particle with $P_s=0$. The effective pumping probability $P_e$ for a real particle, with $P_s\neq0$, is then given by 
\begin{equation}
(P_e|n)=1-(1-P_s)^n.
\label{pe}
\end{equation}
In the limit of $P_s \ll 1$, $(P_e|n)$ is equal to $nP_s$ and thus $P_e = \left<n\right>P_s$, resulting in an upper limit on the value of $P_e$ for real particles, with $\left<n\right>$ representing the maximum factor of improvement offered by the surface patterning.

Appropriate surface structures thus enable increased pumping rate within the same projected area. Labeling the probability that a particle with $P_s=0$ undergoes $n$ collisions as $P(n)$, we can combine the probability distribution for $P(n)$ with (\ref{pe}) to obtain $P_e$ for any value of $P_s$:
\begin{equation}
P_e=\sum_{n=1}^\infty (P_e|n)P(n).    
\label{totprob}
\end{equation}
However, obtaining $P(n)$ is non-trivial and depends on the surface geometry. To do this, we developed a Monte Carlo simulation that models particle incidence upon the structured surface and particle propagation within the structured surface. 

Particles are sampled randomly from an isotropic background gas and propagated, in the long mean free path limit appropriate to high-vacuum environments, until they return through the top plane.

By default we model re-emission of particles that impact the surface using the standard cosine law \cite{cosinelaw}, but we also consider isotropic re-emission into the available half-space and find similar results (see figure \ref{sims}(b)). AM build processes create surface roughness on a scale that is small compared to the intentional feature size but large compared to the length scale of an atom or molecule. In the context of scattering from structured surfaces, this manifests as a partial randomization of the local surface orientation during particle emission, leading to increasingly isotropic emission. By considering the two extreme cases---a pure cosine distribution and fully isotropic emission---we establish bounds within which the actual system performance is expected to fall.

Particles are propagated in this way until they re-emerge from the top-plane of the surface. The number of collisions is then recorded and, by averaging over large numbers of particles ($10^5$ particles per data point for the data shown in figure \ref{sims}(b)), the probability distribution $P(n)$ can be numerically approximated from the results. Equation (\ref{totprob}) can then be applied to determine the increase in capture probability for an incident particle and thus the factor by which the pumping rate is improved, $P_e/P_s$.

\subsection{Simulation results}

\begin{figure*}[ht]
\centering
\includegraphics[width=\linewidth]{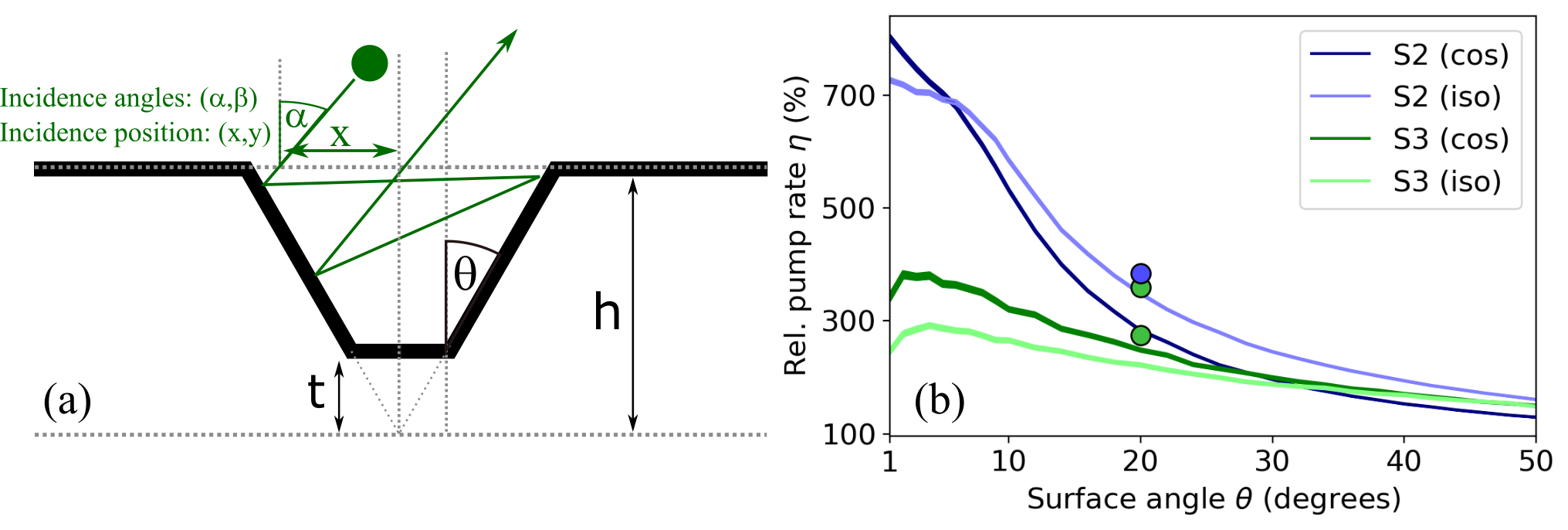}
\caption{(a) Sketch illustrating key simulation parameters; the truncation ratio is given by $t/h$. 
(b) The effect of surface angle $\theta$ on pumping efficiency, for $P_s \ll 1$, for both experimentally tested samples: hexagonal pocket arrays (structure 2) and hexagonally-packed cone arrays (structure 3). Simulations (blue lines for structure 2, green lines for structure 3), for either a cosine-law or isotropic scattering for re-emission of particles from the surface, employed 10$^5$ particles per point. The linewidth represents 1$\sigma$ uncertainty bounds due to numerical error. Circles indicate experimental datapoints for structure 2 (upper point, blue) and upper and lower bounds for structure 3 (middle and lower points, green). Errors on experimental data points are $\pm$10\% (smaller than the marker size). Pockets are assumed fully-tesselated in structure 2.}
\label{sims}
\end{figure*}

\begin{figure*}[ht]
\centering
\includegraphics[width=\linewidth]{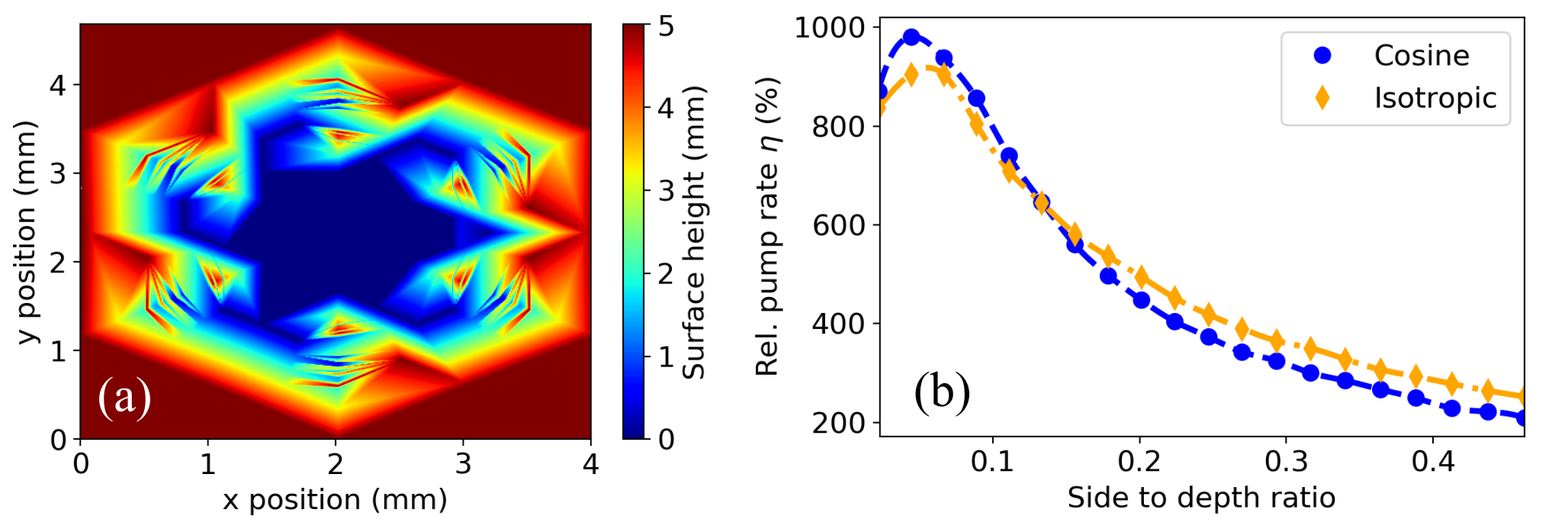}
\caption{(a) Height map of intricately-structured pocket with superior pumping performance. Total depth is set at 5\,mm. (b) Simulated pumping rate increase resulting from the structure displayed in (a) as a function of the ratio between the side length of the hexagon and the total pocket depth, when subject to a uniform stretch along the direction of the normal to the surface top plane. Numerical errors are smaller than the marker size. Dashed lines are cubic interpolations between data. Note that although a direct comparison with simpler geometries is not possible due to the absence of a clearly defined surface angle $\theta$, the \emph{maximum} pumping rate improvement reaches a factor of 9/10 for isotropic/cosine-distributed re-emission, exceeding the respective factors of 7/8 found for polygonal pocket forms.}
\label{escher}
\end{figure*}

Simulations were first conducted for the two surface geometries depicted in figure \ref{samples}: an array of tapered, truncated, hexagonal pockets (structure/sample 2) and a hexagonally-packed array of conical protrusions (structure/sample 3). The exact, realised geometry of the printed structures is shown in the figure. In the case of structure 2, we simulate an idealised form in which the pockets are fully tessellated without interstitial flat regions on the top surface, as such regions occur in our sample only for historical and technical reasons. Figure \ref{sims}(a) depicts the key parameters of the simulation, including the surface angle $\theta$, which is measured between the inclined regions of the structured surface and the normal to the top plane of the surface. Other, more complex forms were also considered and simulated, as illustrated in figure \ref{escher}.  

Our simulations can predict the pumping rate improvement expected for any finite value of $P_s$, but we display here the theoretically optimum case where $P_s \ll 1$. This is a good approximation for many realistic situations \cite{stickprob} and is supported by our experimental results reported in the subsequent sections. 
This approximation inherently becomes a better estimate for the long-term pressure behavior of systems where NEG surface pumping is the main pumping mechanism, since species with low $P_s$ will tend to become the dominant constituents of the residual background gas. 

The simulation results for the geometries in figure \ref{samples} are shown in figure \ref{sims}(b). They indicate that large increases in pumping speed are plausible using fine-scale surface texturing. For example, with $\theta=10^{\circ}$ (an angle that remains plausible for sputter coating \cite{OAD}), fully-tesselated hexagonal pockets are predicted to enhance pumping rate by a factor of between 5.3 and 5.9. They also show the strong dependence of the pumping efficiency on the surface angle $\theta$.  

Figure \ref{sims} shows the results for either a standard cosine-law for re-emission of particles from the surface, or isotropic re-emission into the available half-space. Circles indicate experimental data points for structure 2 (upper point, blue) and upper and lower bounds for structure 3 (middle and lower points, green). The simulations assume complete coverage of an infinite surface. 

Simulation results for $\theta <10^{\circ} $ indicate that the maximum pumping rate increase achievable with these structures is around a factor of 8, achieved for surface angles $\lesssim1^{\circ}$. This is relevant because certain printable materials---such as titanium---can function directly as getter materials when activated at sufficient temperature \cite{tigetter1,tigetter2}, potentially allowing for steeper surface angles in future applications.

In addition, a broad range of patterns were simulated whose cross-sectional shape was that of a regular polygon with 3-8 sides. The relative pumping rate enhancement dependent on $\theta$ is similar for all these shapes with only minor dependence on the number of sides of the polygon and the truncation parameter ($t/h \leq 0.2$). The results are given in the Supplemental Material.

Beyond simple polygonal forms, simulations do predict improvements in performance for some more complex forms. To demonstrate the ability of the simulations to tackle complex surfaces and almost arbitrary geometries, we simulated the effect of intricate pocket shapes based upon geometry-inspired artwork (``Escher tilings''), one of which was found to yield significant improvements in pumping performance. This pocket shape is given by the height map shown in figure \ref{escher}(a), and the corresponding maximum improvement in pumping rate is shown in figure \ref{escher}(b). Because the form simulated involves a range of different surface angles, it cannot be compared directly with the simpler, polygonal pocket forms for any specific surface angle. However, in this case the \emph{maximum} possible pumping rate improvement is considerably higher than that for simple polygonal pocket forms, at a factor 9/10 for isotropic/cosine-distributed scattering respectively. This motivates further investigation of complex, engineered surface patterns for enhancing QT devices and demonstrates the utility of these simulations in analyzing atom-surface interactions.

\section{Sample design and fabrication}

The small-scale surface structures were parametrically designed using the computational tool Grasshopper for Rhinoceros 3D, enabling a scalable approach to the creation of tessellating features for enhanced passive pumping. Although this method was initially applied to planar geometries, it was also found that the method is capable of conformally integrating with more complex, curved surfaces within larger components. This design methodology offers a versatile framework, ensuring that structured surfaces can be seamlessly applied to a wide range of geometrical configurations.

Hexagonal pockets (as recesses) and conical protrusions (as raised features) were selected for exploration due to their efficient packing arrangements, which can easily be implemented in both planar and complex surfaces. The cone structure offers advantages in terms of outgassing and avoiding trapped gas pockets. For the hexagonal pattern, the interconnected nature of the hexagonal pockets enhances the structural integrity during printing, it therefore offers advantages in terms of printability, especially on overhanging features of larger components. Truncation of the structures was employed to avoid build imperfections at sharp features; in the case of recessed features this limits the potential for the formation of trapped gas pockets and thus ensures UHV performance. 

AM enables the creation of sufficiently small and numerous features for compact, integrated application within 3D-printed components (e.g. internal tubes, shields or vacuum parts). Alternatively it permits conformal application over complex internal surfaces of complete vacuum chambers (see Fig.\,\ref{fig:vacuumchamber})\,\cite{AMUHV}. 

While various metals and UHV-compatible glass have been printed successfully \cite{AMUHV,birmingham_flange,"WangCooper2025"}, Ti-6Al-4V Grade 5 is particularly suitable for quantum technology applications, owing to its high strength-to-weight ratio and its relatively low electrical conductivity, which is important in mitigating eddy currents during atom-based sensing experiments. 
All experimental trials were therefore conducted using Ti-6Al-4V Grade 5 and fabricated through laser powder bed fusion. This method provides the necessary precision ($<100~\mu$m resolution) and material properties to realize these designs at the required scale, demonstrating the suitability of additive manufacturing for the production of functional surfaces in advanced applications. 

Following printing and post-processing, the samples (the hexagonal pocket array, hexagonally-packed cone array and a flat reference sample) were sputter coated with V-Zr-Ti NEG material. A composition range was identified from the literature \cite{vzrti} that could be activated at low temperatures around 200°C, which is V~40\%~±~20\%, Zr~40\%~±~20\% and Ti~20\%~±~20\%  (all figures atomic~\% and all intervals absolute~\%~points). 

The sputter coating was performed with a dual DC-magnetron system with one target of V-Zr and one of Ti-V composition, with both magnetrons running simultaneously at appropriate intensity to achieve the desired composition. The system was calibrated to achieve a thickness close to 2~$\mu$m. Energy-dispersive X-ray (EDX) analysis confirmed that the V, Zr and Ti were present in acceptable proportions and that each element was well-dispersed throughout the coating.

\section{Experiment and results}
\label{results}

AM titanium samples were produced as 47.9~mm diameter discs so that, with a machined Conflat\textsuperscript{\textregistered} knife edge, the samples could be incorporated into a standard DN40 rotatable vacuum flange and attached to corresponding vacuum ports. Samples with AM structures 1--3 were produced (see figure \ref{samples}) and NEG coated. An additional flat 3D-printed sample was also manufactured for use as a control sample, without NEG coating.  

Due to initial concerns over potential build imperfections---ultimately revealed to be unnecessary by the quality of the resulting samples---a 500\,$\mu$m flat space was allowed between adjacent pockets in sample 2; this differs from the theoretical optimum version of structure 2 used for simulation purposes, but it is easy to reconcile the two by allowing for the flat area fraction of the real array (see below). 

Visual examination of all samples revealed no significant build errors or distortions and the samples were sputter coated with V-Zr-Ti getter material following the procedure outlined above. 
The entire interior surface of the AM flange, a total area of $A_{R}=1134$\,mm$^2$, was coated with NEG material. The structured surface was restricted to an approximately hexagonal region of total area $A_{S}=726$\,mm$^2$. Within the hexagonal pocket array, flat areas of 500\,$\mu$m width separate adjacent pockets, leaving a remaining $A_P=508$\,mm$^2$ of the area that actually corresponds to the interior of the pockets. 

\begin{figure}[ht]
\centering
\includegraphics[width= \columnwidth]{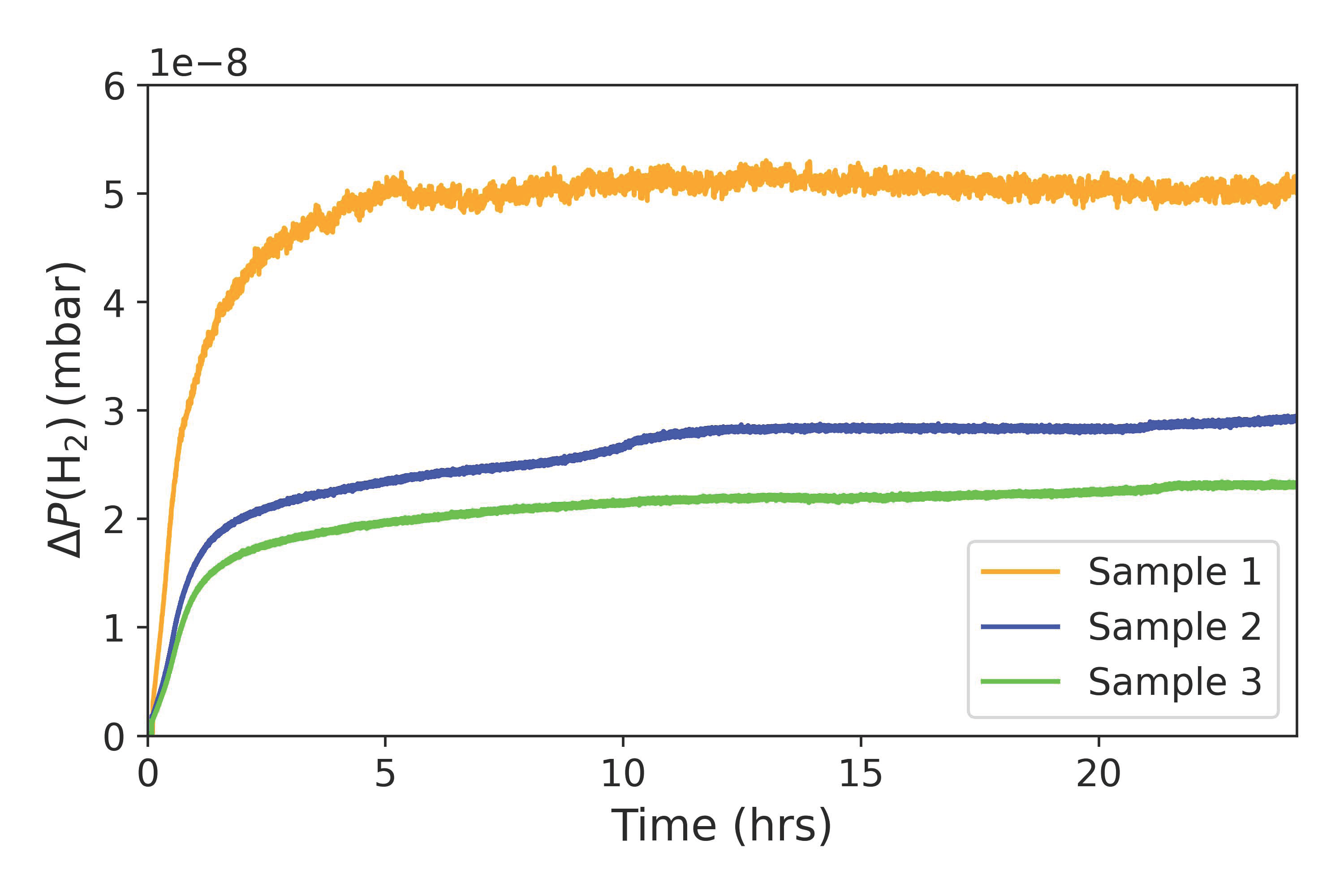}
\caption{Partial pressure increase for H\textsubscript{2} from RGA tests after closing of the gate valve for three different samples. 
The initial partial pressures were $P_\mathrm{H_2}=1.9\times 10^{-10}$\,mbar for sample 1, $P_\mathrm{H_2}=1.2\times 10^{-10}$\,mbar for sample 2 and  $P_\mathrm{H_2}=1.0\times 10^{-10}\,$mbar  for sample 3.}  
\label{torr partial pressure}
\end{figure}

Initial pumping tests were performed using a residual gas analyzer (RGA, Stanford Research Systems RGA100 with electron multiplier), connected to a test rig containing a pressure gauge (Arun Microelectronics AIG17G), 
a turbomolecular pump that could be isolated via an all-metal bonnet valve and the sample under test. Standard baking and activation procedures were applied---see Supplemental Material. 
The bonnet valve was then closed and the changes in partial pressure  for H\textsubscript{2}, starting at the time of valving-off the turbomolecular pump, were recorded by the RGA and are displayed in figure \ref{torr partial pressure}. Partial pressures for CO\textsubscript{2} and N\textsubscript{2}/CO were found to be negligible by comparison; data for this is shown in the Supplemental Material. 
The results clearly confirm that the surface structures of samples 2 and 3 substantially decrease the observed pressure rise, and that hydrogen, which has small $P_s$ for the NEG coating used \cite{stickprob}, is the dominant background gas species. 

\begin{figure*}[ht]
    \centering
    \includegraphics[width=0.9\linewidth]{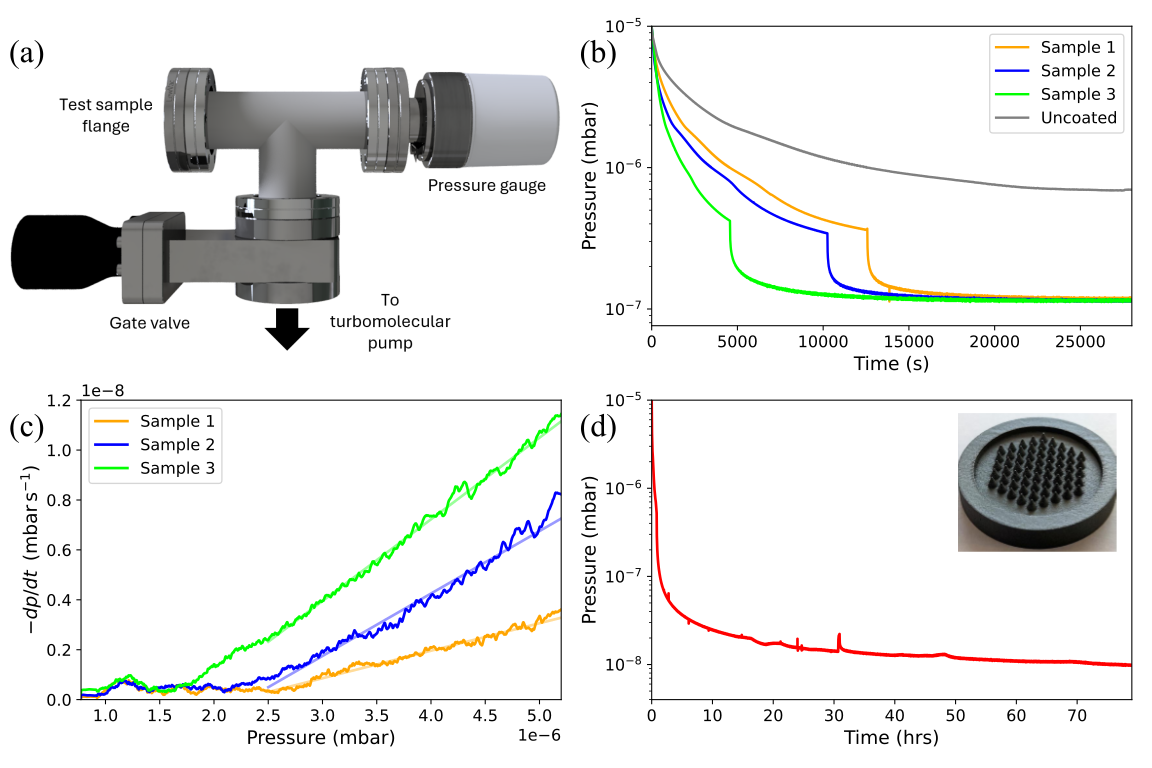}
    \caption{(a) Setup used to test the pumping performance of the additively manufactured and NEG-coated structured samples. The closure of the gate valve allows pumping rates to be inferred from pressure data recorded by the vacuum gauge. (b) Measured chamber pressure against time for each test sample including a printed, uncoated sample. 
    (c) Net pumping rates $- dp/ dt$ for each NEG coated sample as a function of pressure. Linear fits have been used to calculate the pumping rate coefficients for the pumped volume, which are $\gamma_1=1.10(2)\times10^{-3}\,\mathrm{s}^{-1}$, $\gamma_2=2.50(3)\times10^{-3}\,\mathrm{s}^{-1}$ and $\gamma_3=3.27(3)\times 10^{-3}\,\mathrm{s}^{-1}$. 
    (d) Long term pressure data for sample 3 (pictured) showing that pumping performance is maintained over $>3$ days and a final pressure of $9.7\times 10^{-9}$ mbar is reached.}
    \label{mainres_uon}
\end{figure*}

To quantitatively assess the pumping performance increase, tests were conducted using the apparatus shown in figure \ref{mainres_uon}(a). The chamber, with an internal volume of  0.20\,l, was evacuated and baked, and the NEG-coating activated as described in the Supplemental Material. 

The angle valve seen in figure \ref{mainres_uon}(a) was then closed and the pressure as a function of time monitored using the pressure gauge (Pfeiffer IKR 270 Compact Cold Cathode Gauge). An initial pressure spike occurred due to the disturbance caused by physically closing the valve. This was followed in each case by a gradual decline in the measured pressure. The raw data is shown in figure \ref{mainres_uon}(b). A small decrease in pressure is also observed for an uncoated sample (grey line in Fig.~\ref{mainres_uon}(b)). This can most likely be explained by contamination of the system with NEG material from previous tests, or by a pumping contribution from the ion gauge. By performing a differential measurement, comparing structured NEG-coated samples to a flat NEG-coated sample, the influence of all such sources of spurious pumping or outgassing can be negated. 

In order to determine a pumping rate coefficient $\gamma$ (see equation (\ref{pressuredrop}) below) under comparable conditions, each data set was truncated at the point at which the measured pressure reached 10$^{-5}$ mbar during the decline following the initial spike; this ensured repeatable and consistent starting conditions. 
The rate of pressure decrease $dp/dt$ was then determined in each case as a function of pressure, using linear interpolation, where necessary, to enable a consistent set of data points across all samples. 
To eliminate the influence of spurious pumping/contamination sources, the rate of pressure drop for the uncoated sample was subtracted. 

The residual rate of pressure drop attributable to the NEG coating was then plotted against pressure in each case---see figure \ref{mainres_uon}(c). The net rate of pressure drop under constant pumping conditions is given by
\begin{equation}
\frac{dp}{dt} = C - \gamma p,
\label{pressuredrop}
\end{equation}
where $C$ is a constant rate of pressure increase resulting from leaks and outgassing, $p$ is the current pressure in the system and $\gamma$ is the pumping rate coefficient, which represents the `strength of pumping' \cite{NEGpumpingrate, Gaugepumpingrate}. 

In figure \ref{mainres_uon}(c), $-dp/dt$ is plotted against $p$ and the gradient corresponds to the pumping rate coefficient $\gamma_i$ for each sample. The ratio of the fitted gradients for the different samples therefore gives the ratio of the pumping rate coefficients of the different surfaces. 

The results are summarized in table \ref{Table:pumping constants} and show that sample 3 offered an NEG-based pumping rate coefficient greater than that of sample 1 by a factor of $\gamma_{3}/ \gamma_{1} = 3.0 \pm 0.1$, while structure 2 increased it by a factor of $\gamma_2/\gamma_1=2.3 \pm 0.1$.

In order to quantify the improvement introduced by the 3D-printed geometry and to compare to simulations, a useful measure is the pumping rate coefficient per unit projected area of the structured surface, denoted $\gamma_{i,s}$ for sample $i$.

To calculate $\gamma_{i,s}$, allowance must be made for the flat parts of the sample, as well as for the fact that particles can enter through the vertical sides of the array in structure 3 (see Supplemental Material). This second consideration results in the determination of upper and lower bounds to $\gamma_{3,s}$, depending on what assumptions are made about pumping by the sides of the array.

These calculations yield enhancements factors,  $\eta_i=\gamma_{i,s}/\gamma_{1,s}$, of the pumping rate coefficient per unit area given by $\eta_2=3.0 \pm 0.1$ and upper and lower limits of $\eta_{3,\mathrm{max}}=3.6 \pm 0.1$ and $\eta_{3,\mathrm{min}}=2.7 \pm 0.1$ for structures 2 and 3 respectively. These factors of improvement show that the method offers substantial benefits even with relatively conservative surface designs. It is interesting to note that $\eta$ exceeds the ratio by which the total surface area is increased. 

For the purposes of comparison with simulation, $\eta_{2^*}$---the improvement in pumping rate coefficient per unit projected area for the hexagonal pockets themselves, excluding the interstitial flat regions---is found to be $3.8 \pm 0.1$. The results are displayed in figure \ref{sims} and include both bounds for structure 3. 

Simulation and experiment show good correspondence, with the experimental data displaying a slightly larger pumping rate enhancement than expected. This can be explained by the influence of small-scale surface roughness caused by the AM build process, which was not included in the simulation. 

Overall, the experimental results confirm the basic premiss of our theory and show that it provides a good description of actual pumping performance. In particular, our simulations are suitable estimates to predict the performance of AM surface structures of known geometry. This further supports the simulation results shown in figures \ref{sims} and \ref{escher}, which indicate that much larger improvements in pumping performance are possible.
For example, figure \ref{escher} predicts $7.5<\eta<8$ for pockets of the form shown with a side to depth ratio of 0.1; given that standard laser powder bed fusion manufacturing offers a resolution  $<100\,\mu$m, this can be achieved with absolute pocket depths of a few mm. At the optimal side to depth ratio of 0.028, simulations predict $9.1<\eta<9.8$ for this pocket shape.    

Figure \ref{mainres_uon}(d) shows longer-term pressure data taken with sample 3, indicating that this sample can pump a volume of 0.20\,l for $>70$ hours of operation to a pressure $<10^{-8}$\,mbar. As no degradation of performance was observed within the window of observation, this data provides only a lower limit on pump longevity, which may in fact be far longer. 
In general, the increase in total surface area when applying a NEG coating to a structured surface, rather than a flat one, increases the total pump capacity and thus extends operational longevity. \\

\begin{table}
\centering
\setlength{\tabcolsep}{5.5pt} 
\renewcommand{\arraystretch}{1.3} 
\begin{tabular}{ |c| c c c|} 
 \hline
 AM sample & Rel. SA  & $\gamma_i/\gamma_1$ & $\eta_i$ \\ 
 \hline\hline
 1 & 1 & $1$ & 1 \\ 
 \hline
 2 & 2.3 & $2.3(1)$ & $3.0(1)$ \\
 2$^*$ & 2.9 & NA & $3.8(1)^{*}$ \\
 \hline
\multirow{2}*{3} & \multirow{2}*{1.9} & \multirow{2}*{$3.0(1)$} & $2.7(1)$ (min) \\

 & & & $3.6(1)$ (max)\\
 \hline
\end{tabular}
\caption{Comparison of the experimentally tested samples; relative total surface area per unit projected area of the structured surface, Rel. SA, measured pumping rate coefficient enhancement $\gamma_i/\gamma_1$ for each sample flange and corresponding relative pumping rate coefficient per unit structured area, $\eta_i=\gamma_{i,s}/\gamma_{1,s}$, for each sample. Sample 2$^*$, together with the related $\eta_2^{*}$ inferred from the measured $\gamma_2$, represents the theoretical optimum form of sample 2 without interstitial flat regions.} 
\label{Table:pumping constants}
\end{table}

\section{Conclusions}

We have demonstrated that additively manufactured, complex surface structures can be employed to manipulate gases in ultra-high vacuum chambers. 
In particular, we have shown that structures 3D-printed in $^{64}$Ti are UHV compatible and enable a tailored enhancement in the number of surface scattering events induced by the structured elements. Outgassing tests (see Supplemental Material) did not measure any adverse effects of the structure or build process on vacuum performance.
Applications include controlling particle propagation pathways and dwell times in UHV devices, surface pumping with high efficiency. 
Experimentally, we have shown a 3.8-fold increase in pumping rate per unit area by adding a purpose-designed surface texture to an NEG-coated surface. We have developed and presented Monte Carlo simulations that enable the prediction of the performance of complex AM surface geometries. Comparing theory and experiment, we find good agreement, while the experimental results slightly exceed the predictions of Monte Carlo simulations. 
These simulations reveal a range of interesting surface structures that offer even greater improvements in pumping rate, predicting up to a 10-fold increase in pumping rate for some more complex forms. 

Promising directions for further development include adapting the technique to provide optimum performance from components that control gas flow via impact with a surface, such as collimation nozzles \cite{dual_oven,csovens} and cold plates \cite{BEC_machines}. Conformal application over a large fraction of a vacuum system's interior surface, coupled to direct activation of the titanium build material as an NEG pump, is also a very interesting avenue of future research, with the potential to enable extreme pumping performance and underpin portable UHV and even XHV systems. 

The capabilities provided by such engineered AM surface structures are relevant to numerous technologies and devices across QT and beyond. QT applications will include next-generation portable gravimeters \cite{PortableCAG,PRAp_grav}, cold-atom magnetometers \cite{rfmagnetometer,PRAp_mag2}, atomic clocks \cite{portableatomicclock,PRAp_clock2} and quantum accelerometers \cite{Atomacceleratometer}. Application areas outside QT include mass spectrometers \cite{C8JA00087E}, particle accelerators \cite{linacc} and electron microscopes, especially where the use of methods such as environmental transmission electron microscopy or the study of novel materials imposes additional pumping and gas-flow requirements \cite{TEM,ETEM}. The passive nature of such surfaces and the ease with which AM can shape components for integration into compact, lightweight systems makes this approach ideally suited to field-deployable \cite{Qenergy} or space-based \cite{qtinspace} devices.

\section{Acknowledgements}
This work has been supported by the UK Engineering and Physical Sciences Research council (EPSRC) grants EP/T001046/1, EP/R024111/1, EP/Y005139/1,  EP/Z533166/1, the Innovate UK projects 10143484 (QAPTCHA) and 10031462 (QTEAM), the grant 62420 from the John Templeton Foundation, and the EPSRC Impact Acceleration Account EP/X525765/1.

\bibliography{sample}


\setcounter{equation}{0}
\renewcommand\theequation{S\arabic{equation}}
\setcounter{figure}{0}
\renewcommand{\thefigure}{S\arabic{figure}}

\bigskip
\bigskip

\appendix

\noindent \textbf{\large Supplementary Material}
\bigskip




\noindent \textbf{Simulation procedure and full results}
\smallskip

Pumping simulations were conducted in Python using a Monte-Carlo method. The particles in the main volume of the vacuum system were assumed to be of homogeneous density and have a Maxwell-Boltzmann distribution of velocities, such that the probability density for a particle to enter through the top plane of the structured surface is independent of its position on the surface and proportional to the cosine of the angle between its direction of propagation and the normal to the top plane of the surface. The temperature of the particles is irrelevant as only the direction of particle incidence, not their speed, affects their propagation route within the structured surface (and as such any isotropic distribution of velocities would yield the same results). The pressure was assumed to be sufficiently low that the mean free path of the particles is long compared to the scale of the surface features, while the temperature of the background gas and surface is taken to be high enough that gravity can be ignored when determining particle trajectories within the structured surface. 

Particles were thus propagated along straight lines until they impacted the surface. At the point of impact they were taken to be re-emitted with a probability distribution for their direction of emission given according to either the cosine rule (probability density per unit solid angle proportional to the cosine of the angle between the emission direction and the local surface normal) or an isotropic distribution within the available half-space. They were then propagated again until they impacted either the surface or its top plane, and each time they impacted the surface this process was repeated. When a particle was incident upon the top plane of the surface, such that it would pass through and leave the structured surface into the main volume of the vacuum system, the total number of surface collisions it had undergone was recorded. 

A large number of particles were simulated sequentially in each case, and the fraction of particles undergoing each possible number of collisions was used to determine $P(n)$ for a given surface. 

Supplementary figure 1 shows the effects of pocket truncation in panel (a) and of pocket shape in panel (b). The results show that small truncations have very little effect on pumping performance, though larger degrees of truncation do significantly reduce performance, as expected. Panel (b) clearly shows that the influence of pocket shape is small, with the key parameters being the face inclination and truncation ratio. Note that the enhancement factors displayed in the figure assume that the particles are incident within such a pocket --- for non-tessellating shapes it will not be possible to achieve as large a level of improvement, since there will necessarily be some flat regions between the pockets on any continuous surface. 

\begin{figure*}[ht]
\centering
\includegraphics[width=0.9\linewidth]{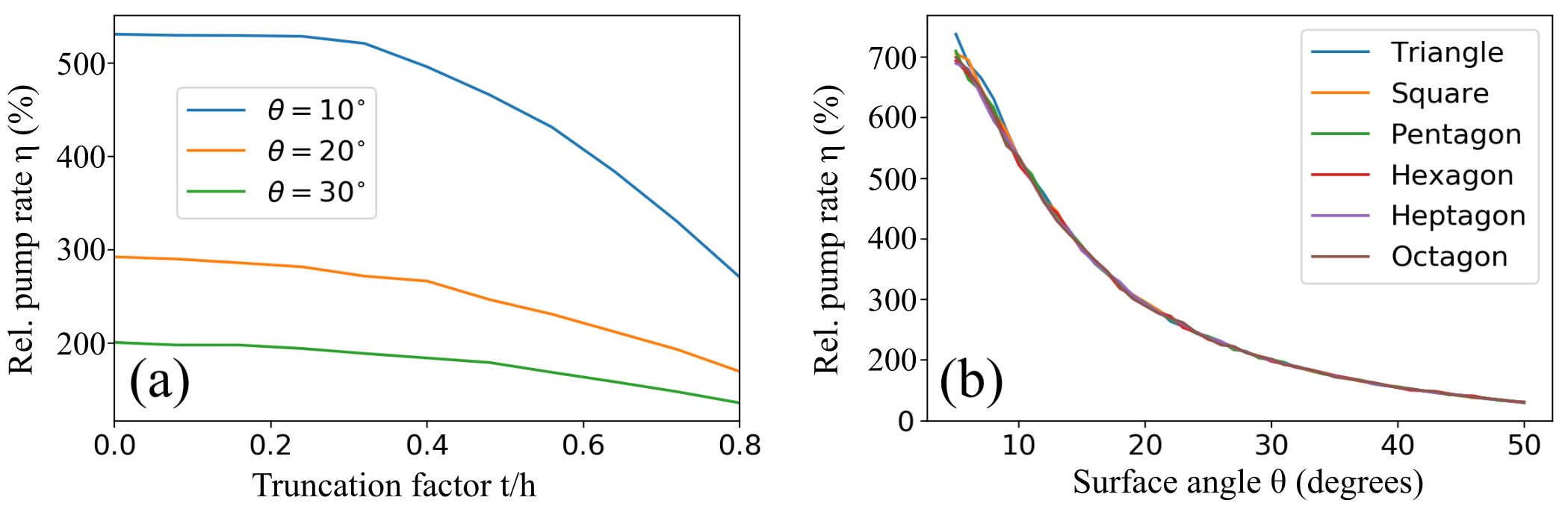}
\caption{(a) Simulated pumping efficiency of truncated hexagonal pockets as a function of truncation scale ratio (as defined in figure 3 of the main article), for surface angles of 10, 20 and 30 degrees. (b) Simulated pumping efficiency of truncated pockets whose cross sections are those of regular polygons with 3 to 8 sides, at a truncation ratio of 0.2, as a function of surface angle $\theta$. Both panels assume a cosine law re-emission pattern.} 
\label{shapetrunc}
\end{figure*}

In figure 4 of the main article, we present results for a complex pocket shape based on geometric artwork. Results for a similar but yet more complex pocket shape are presented in supplementary figure \ref{esch2} below.

\begin{figure*}[ht]
\centering
\includegraphics[width=0.9\linewidth]{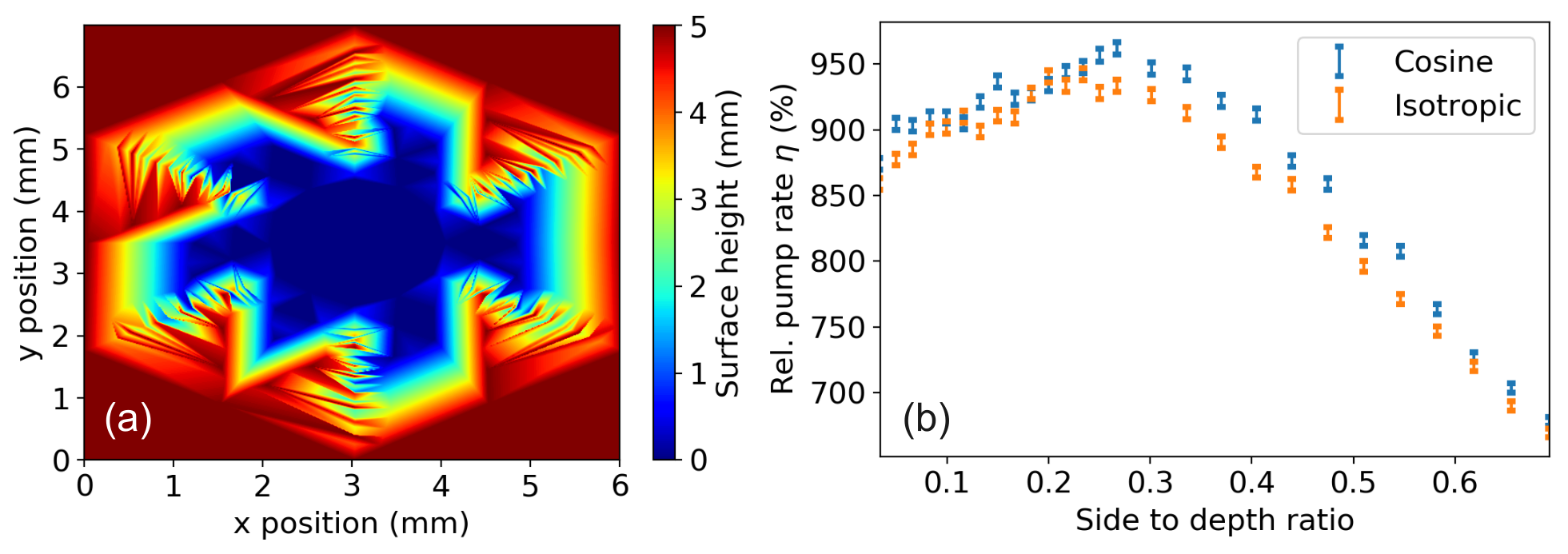}
\caption{(a) Height map showing the exact structure of a complex pocket shape based on geometric artwork, for a total pocket depth of 5\,mm. (b) Simulated pumping efficiency increase for this pocket shape, as a function of the ratio between the side length of the outer hexagon and the depth of the pocket, when subject to uniform stretch along the direction of the normal to the surface top plane. Simulations use 3.5$\times 10^4$ particles per point. Performance is slightly below that for the similar structure described in the main article, with the broadening of the response as a function of side to depth ratio linked to the smaller size of the individual features within the pocket compared to the overall pocket size.} 
\label{esch2}
\end{figure*}

\bigskip

\noindent \textbf{Baking and activation procedures}
\smallskip

For the data taken using the residual gas analyzer (RGA) and displayed in figure 5 of the main article, the system was baked to 200°C for 48 hours and allowed to cool, following which the RGA and pressure gauge filaments were degassed. The NEG material was activated by local heating to 250°C for 48 hours and allowed to cool.

For the second set of data collection, which provided the data shown in figure 6 of the main article, the chamber was evacuated using a turbomolecular pump and baked at $180^\circ$C for 3 days. Following this, the sample flange temperature was increased to $300^\circ$C and the rest of the apparatus to $270^\circ$C over eight hours; this temperature was then maintained for a further 60 hours. Following this, the heating was deactivated and the chamber allowed to cool for 16 hours in an environmental temperature of $21^\circ$C.

\bigskip

\noindent \textbf{Vacuum compatibility tests for uncoated \\ samples}
\smallskip

To address the question of potential outgassing of the printed material, additional vacuum testing was conducted with as-printed samples, without NEG coating using a residual gas analyzer (RGA, Stanford Research Systems RGA100 with electron multiplier). Flat samples of varying thickness (1, 3 and 6\,mm) were tested. All samples were vacuum baked and solvent cleaned prior to mounting. The test rig was baked under vacuum to $200^\circ$~C for 48 hours then allowed to cool. The filaments of the RGA and vacuum gauge (Arun Microelectronics AIG17G gauge) were degassed. The test setup was similar to the one shown in figure 6 of the main article, with a turbomolecular pump that can be isolated by closing a gate valve. After pumping down to base pressure the turbomolecular pump was valved off and the subsequent rise in pressure was recorded. Invariably, the closing of the valve caused a spike in pressure which quickly subsided. A standard 304L stainless steel sample (machined, not printed) of thickness 6~mm was tested as a baseline comparison. \\
The increase of pressure following the valving off of the turbomolecular pump allowed calculation of outgassing rates for the chamber as a whole with the sample in place. For the flat samples tested, measured outgassing rates were $3.6\times10^{-13}$\,mbar\,L\,s\textsuperscript{-1}\,cm\textsuperscript{-2} (1\,mm thick), $1.8\times 10^{-13}$\,mbar\,L\,s\textsuperscript{-1}\,cm\textsuperscript{-2} (3\,mm thick) and $1.1\times 10^{-13}$\,mbar\,L\,s\textsuperscript{-1}\,cm\textsuperscript{-2} (6~mm thick). In these cases the outgassing rate was less than or the same as the 6 mm thick stainless steel baseline sample ($1.1\times10^{-12}$\, mbar\,L\,s\textsuperscript{-1}\,cm\textsuperscript{-2}).

\bigskip

\noindent \textbf{RGA data for NEG-coated samples}
\smallskip

RGA measurements following valving-off of the turbopump reveal equivalent data to that shown for hydgrogen in figure 5 of the main article for CO$_2$ and CO/N$_2$. This is shown in supplementary figure 2. 
The samples are the same as introduced in the main text, sample 1 - flat, sample 2 - hexagonal pockets, sample 3 - conical protrusions. 
Note, that the partial pressures of these species are around two orders of magnitude lower than that for hydrogen, and their contribution to the overall pressure is therefore negligible. 

\begin{figure*}[ht]
\centering
\includegraphics[width=0.9\linewidth]{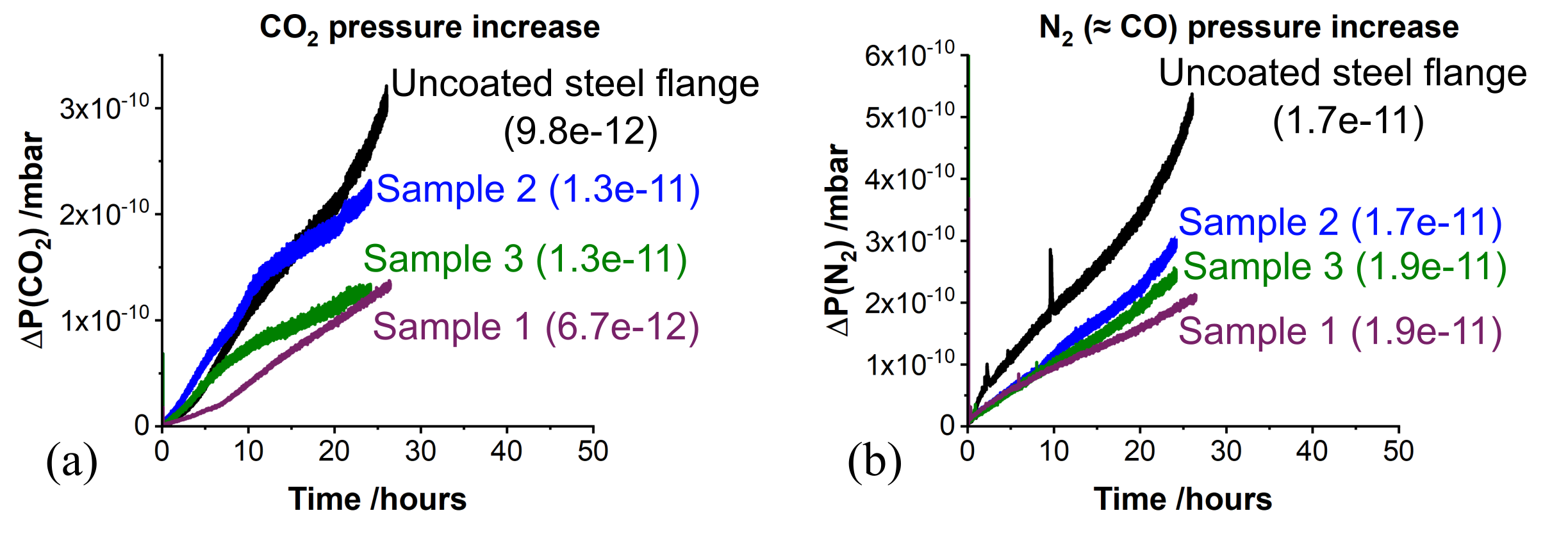}
\caption{Pressure rise data following valving-off of the turbopump on the RGA setup for CO$_2$ (a) and CO/N$_2$ (b). The numbers in brackets indicate the initial pressure for each case at $t=0$, in mbar. The ``$\approx$" symbol for N$_2$ and CO in panel (b) indicates that the analysis performed cannot distinguish between these molecules, and the pressure values given are the total across both species.} 
\label{rga_test_setup}
\end{figure*}

\bigskip

\noindent \textbf{Calculation of $\gamma_{i,s}$}
\smallskip

With flat area $F_A$ and structured area $S_A$ present in sample $i$, one finds that
\begin{equation}
    \gamma_{i,s}=\frac{\gamma_i - \gamma_{1,s}F_A}{S_A}, 
\end{equation}
where $\gamma_{1,s}$ is simply equal to the pumping rate coefficient for sample 1 divided by the total flat, coated area, such that $\gamma_{1,s}=\gamma_1/A_R$. 

In the case of sample 2 $F_A=A_R-A_S$ while $S_A=A_S$. To obtain $\gamma_{2^*,s}$, the pumping rate coefficient per unit area for the theoretically optimum form of sample 2 with no interstitial flat regions, as considered in the simulations, one uses $F_A=A_R-A_P$ and $S_A=A_P$.

In the case of sample 3, the real, finite array allows particles to enter through the boundary of the array \emph{below} the top plane of the surface (at the sides of the array), increasing the pumping rate beyond that attributable only to incidence through the top plane of the structure. This does not match the simulations, which required wrapped boundary conditions to handle the structure's open unit cell and therefore simulated an infinite surface. Nor does it match many typical deployment conditions, in which the surface covered may be large enough to render the influence of pumping via the sides of the array negligible. The measured pumping rate must therefore be scaled correctly to obtain the pumping performance of a surface of large extent. It is not trivial to do this, as incidence onto the sides of the array cannot be simulated using only a repeated unit cell. We therefore place a lower limit on the pumping rate enhancement for an infinite surface where this effect cannot take place by assuming that the entire area of the sides of the structure ($A_V=354\,\mathrm{mm}^2$) constitutes a normal part of the structured surface for the purposes of our experimental measurements. This gives $F_A=A_R-A_S$ while $S_A=A_S+A_V$. An upper bound is found by assuming that the sides of the structure pump as effectively as a flat surface of equal area, giving $F_A=A_R-A_S+A_V$ and $S_A=A_S$.

\typeout{get arXiv to do 4 passes: Label(s) may have changed. Rerun}

\end{document}